\documentstyle[11pt,fleqn]{article}
\catcode`\@=11
\@addtoreset{equation}{section}
\def\theequation{\arabic{section}.\arabic{equation}}
\def\appendix{\renewcommand{\thesection}{\Alph{section}}\setcounter{section}{0}
              \renewcommand{\theequation}
            {\mbox{\Alph{section}.\arabic{equation}}}\setcounter{equation}{0}}
\def\maketitle{\thispagestyle{empty}\setcounter{page}0\newpage
                \renewcommand{\thefootnote}{\arabic{footnote}}
                  \setcounter{footnote}0}
\renewcommand{\thanks}[1]{\renewcommand{\thefootnote}{\fnsymbol{footnote}}
               \footnote{#1}\renewcommand{\thefootnote}{\arabic{footnote}}}
\newcommand{\preprint}[1]{\hfill{\sl preprint - #1}\par\bigskip\par\rm}
\renewcommand{\title}[1]{\begin{center}\Large\bf #1\end{center}\rm\par\bigskip}
\renewcommand{\author}[1]{\begin{center}\Large #1\end{center}}
\newcommand{\address}[1]{\begin{center}\large #1\end{center}}
\def\dip{\smallskip Department of Mathematics, University of Massachusetts,\\
           Amherst, Massachusetts 01003}
\def\Idip{\address{\dip}}
\newcommand{\email}[1]{e-mail: \sl #1@math.umass.edu\rm}
\newcommand{\femail}[1]{\thanks{\email{#1}}}

\def\babs{\hrule\par\begin{description}\item{Abstract: }\it}
\def\eabs{\par\end{description}\hrule\par\medskip\rm}
\renewcommand{\date}[1]{\par\bigskip\par\sl\hfill #1\par\medskip\par\rm}
\newcommand{\ack}[1]{\par\section*{Acknowledgments} #1}
\newcommand{\s}[1]{\section{#1}}
\renewcommand{\ss}[1]{\subsection{#1}}

\renewcommand{\vec}[1]{{\bf #1}}       
\def\beq{\begin{eqnarray}}    
\def\eeq{\end{eqnarray}}      
\newtheorem{theorem}{Theorem}                  
\newtheorem{proposition}{Proposition}          
\newtheorem{remark}{Remark}                    
\def\R{{\hbox{{\rm I}\kern-.2em\hbox{\rm R}}}}   
\def\H{{\hbox{{\rm I}\kern-.2em\hbox{\rm H}}}}   
\def\N{{\hbox{{\rm I}\kern-.2em\hbox{\rm N}}}}   
\def\C{{\ \hbox{{\rm I}\kern-.6em\hbox{\bf C}}}} 
\def\Z{{\hbox{{\rm Z}\kern-.4em\hbox{\rm Z}}}}   
\renewcommand{\Re}{\mathop{\rm Re}\nolimits}       
\def\al{\alpha}
\def\be{\beta}
\def\ga{\gamma}

\def\Ga{\Gamma}

\def\citen#1{%
\edef\@tempa{\@ignspaftercomma,#1, \@end, }
\edef\@tempa{\expandafter\@ignendcommas\@tempa\@end}%
\if@filesw \immediate \write \@auxout {\string \citation {\@tempa}}\fi
\@tempcntb\m@ne \let\@h@ld\relax \let\@citea\@empty
\@for \@citeb:=\@tempa\do {\@cmpresscites}%
\@h@ld}
\def\@ignspaftercomma#1, {\ifx\@end#1\@empty\else
   #1,\expandafter\@ignspaftercomma\fi}
\def\@ignendcommas,#1,\@end{#1}
\def\@cmpresscites{%
 \expandafter\let \expandafter\@B@citeB \csname b@\@citeb \endcsname
 \ifx\@B@citeB\relax 
    \@h@ld\@citea\@tempcntb\m@ne{\bf ?}%
    \@warning {Citation `\@citeb ' on page \thepage \space undefined}%
 \else
    \@tempcnta\@tempcntb \advance\@tempcnta\@ne
    \setbox\z@\hbox\bgroup 
    \ifnum\z@<0\@B@citeB \relax
       \egroup \@tempcntb\@B@citeB \relax
       \else \egroup \@tempcntb\m@ne \fi
    \ifnum\@tempcnta=\@tempcntb 
       \ifx\@h@ld\relax 
          \edef \@h@ld{\@citea\@B@citeB}%
       \else 
          \edef\@h@ld{\hbox{--}\penalty\@highpenalty \@B@citeB}%
       \fi
    \else   
       \@h@ld \@citea \@B@citeB \let\@h@ld\relax
 \fi\fi%
 \let\@citea\@citepunct
}
\def\@citepunct{,\penalty\@highpenalty\hskip.13em plus.1em minus.1em}%
\def\@citex[#1]#2{\@cite{\citen{#2}}{#1}}%
\def\@cite#1#2{\leavevmode\unskip
  \ifnum\lastpenalty=\z@ \penalty\@highpenalty \fi 
  \ [{\multiply\@highpenalty 3 #1
      \if@tempswa,\penalty\@highpenalty\ #2\fi 
    }]\spacefactor\@m}
\begin{document}

\preprint{}

\title{Product Einstein Manifolds, Zeta-Function Regularization and the
Multiplicative Anomaly}

\author{Andrei A. Bytsenko\thanks{email: abyts@spin.hop.stu.neva.ru}}
\address{State Technical University, St. Petersburg 195251, Russia}
\author{Floyd L. Williams\femail{williams}}
\Idip

\date{June 1997}

\babs
The global additive and multiplicative properties of Laplace type operators
acting on irreducible rank 1 symmetric spaces are considered. The explicit
form of the zeta function on product spaces and of the multiplicative anomaly
is derived. 

\eabs

\s{Introduction}

Among the most important geometric structures on manifolds
\cite{yano65,cher66-72-167,koba72} almost-product structures play an essential
role. Structures of this kind appear in a natural way from a
variational principle based on a general class of Lagrangians depending on the
Ricci square invariant constructed out of a metric and a symmetric connection
\cite{boro96u-67,boro96u}. Geometric properties of (pseudo-) Riemannian
almost-product structures have been studied in Refs. \cite{yano65,rein83,
nave83-3-577,gil83-26-358,roca88-32-654}.

In this class of structures an almost-product structure on Einstein manifold
is associated with an Einstein metric ${\bf g}$ (i.e. $\mbox{Ric}({\bf g})=
\ga{\bf g}$ holds,
where $\mbox{Ric}({\bf g})$ is the Ricci tensor and $\ga$ is a constant).
It is well known that gravitational Lagrangians which are non-linear in the
scalar curvature of a metric give rise to equations with higher derivatives or
to appearance of additional matter fields. An important example of a non-linear
Lagrangian leading to equations with higher derivatives is given by Calabi's
variational principle \cite{cala82-102-259}. The non-linear gravitational
Lagrangians which still generate Einstein equations are particularly important
since they provide a general approach to governing topology in low dimension
models and can be adapted in string theory.

In higher derivative field theories (for example, in higher derivative quantum
gravity) as a consequence one has to deal with the product of two (or more)
elliptic differential operators \cite{birr82,buch92}. In some cases an elliptic
(pseudo-) differential operator has a well-defined zeta-regularized determinant.
Is is natural to investigate multiplicative properties of the determinants of
differential operators, in particular the so-called multiplicative anomaly
\cite{kont94u-56,kont94u-40,eliz97u-60}; for the definition of the anomaly see
Sect. 3 below. The multiplicative anomaly can be expressed by means
of the non-commutative residue associated with a classical pseudo-differential
operator, the Wodzicki residue \cite{wodz87}.

Recently the important role of this residue has been recognized in physics.
The Wodzicki residue, which is the unique extension of the Dixmier trace
to the wider class of pseudo-differential operators (PDOs)
\cite{conn88-117-673,kast95-166-633}, has been considered within the
non-commutative geometrical approach to the standard model of the electroweak
interactions \cite{conn90-18-29,conn94,conn96-53,cham96-01, cham96-56,
mart96-01}. This residue is also used to write down the Yang-Mills action
functional. Other recent papers along these lines can be found in Refs.
\cite{kala95-16-327,acke95-06,acke95-52}. The residue formulae have been used
also for dealing with the singularity structure of zeta functions
\cite{eliz96-56} and the commutator anomalies of current algebras
\cite{mick94-93}.

The purpose of the present paper is to investigate the global additive and
multiplicative properties of invertible elliptic operators of Laplace type
acting on Einstein manifolds, especially on a non-compact symmetric space,
and related zeta functions. The contents
of the paper are the following. In Sect. 2 we summarize some properties of
the spectral zeta function $\zeta(s|\bigoplus_{p}{\cal L}_p)$ related to the
Laplace type operators ${\cal L}_p,\,\,\,p=1,2$, each of them acting on an
irreducible rank 1 symmetric space of non-compact type. The explicit form of
the zeta function is given for a wide class of symmetric space products, in
particular for the simplest case $H^2\otimes H^2$. In Sect. 3 the spectral
zeta function $\zeta(s|\bigotimes_{p}{\cal L}_p)$ and the corresponding
multiplicative anomaly (for the spaces mentioned above) are studied.
The explicit calculation is given for essentially all rank 1 symmetric spaces.
We end with some conclusions in Sect. 4. Finally the Appendices A and B
contain a summary of a meromorphic continuation of the zeta functions
$\zeta(s|\bigoplus_{p}{\cal L}_p)$ (Appendix A) and
$\zeta(s|\bigotimes_{p}{\cal L}_p)$ (Appendix B).

\s{Determinant Regularization and Product of Einstein Manifolds}

In this section we consider the problem of the global existence of zeta
functions on (pseudo-) Riemannian product manifolds, a product of two
Einstein manifolds \cite{yano65,yano84}
$$
(M,{\bf g},{\cal P})=(M_1,{\bf g}_1,{\cal P}_1)\otimes (M_2,{\bf g}_2,
{\cal P}_2)\mbox{,}
\eqno{(2.1)}
$$
where ${\bf g}={\bf g}_1\otimes{\bf g}_2$ and the metric ${\bf g}$ separates
the variables, i.e.
$$
 ds^2={\bf g}_{\al\be}(x)dx^\al\otimes dx^\be+{\bf g}_{\mu\nu}(y)dy^\mu
\otimes dy^\nu\mbox{.}
\eqno{(2.2)}
$$
The tangent bundle splits as $TM=TM_1\oplus TM_2$ and ${\cal P}={\cal P}_1+
{\cal P}_2$, where ${\cal P}_p\,\,(p=1,2)$ are the corresponding projections
on $TM_p$,
$$
\mbox{Ric}({\bf g})=\ga{\bf g}\mbox{,}
\eqno{(2.3)}
$$
$$
{\cal P}^2=Id,\,\,\,\,\,\,{\bf g}({\cal P}{\cal X},{\cal P}{\cal Y})={\bf g}
({\cal X},{\cal Y}),\,\,\,\,\,\,
{\cal X},{\cal Y}\in
\chi(M)\mbox{,}
\eqno{(2.4)}
$$
and $\chi(M)$ being the Lie algebra of vector fields ${\cal X}$ and ${\cal Y}$
on $M$. The trivial examples of an almost-product structure are given by the
choices ${\cal P}=\pm Id$ ($\pm$ identity).

We recall some facts about Einstein manifolds necessary for the next
considerations.
We start with an almost-product (pseudo-) Riemannian structure
$(M,{\bf g},{\cal P})$ which is integrable iff $\triangle_{\bf g}{\cal P}=0$ for
the Levi-Civita connection $\triangle_{\bf g}$ of ${\bf g}$. The two integrable
complementary subbundles, i.e. both foliations, are totally geodesic
\cite{yano65,yano84}. Let $M$ be a pseudo-K{\"a}hler manifold; the metric
${\bf g}$ splits as ${\bf g}={\bf g}_1\otimes{\bf g}_2$. Such a manifold is
an Einstein manifold iff in any adapted co{\"o}rdinates $(x^\alpha,y^\alpha)$ both
metrics ${\bf g}_1$ and ${\bf g}_2$ are Einstein metrics for the same constant
$\ga$ \cite{yano65,yano84,bess87}, i.e.
$$
R_{\alpha_{p}\beta_{p}}=\ga g_{\alpha_{p}\beta_{p}},\,\,\,\,\,(p=1,2)
\mbox{.}
\eqno{(2.5)}
$$

Our consideration will be restricted to only locally decomposable manifolds.
A wide class of (pseudo-) Riemannian manifolds includes non-locally
decomposable manifolds as well, which are given by warped product space-times
\cite{onei83,desz91-23-671,caro93-10-461}. Note that many exact solutions of
Einstein equations (associated with Schwarzschild, Robertson-Walker, Reissner-
Nordstr{\"o}m, de Sitter space-times) and $p$-brane solutions
\cite{aref96u-25,boro96u} are, in fact, warped product space-times
\cite{bess87,caro93-10-461}.

\ss{The Spectral Zeta Function}

We shall be working with irreducible rank 1 symmetric spaces $M_p\equiv X_p=
G_p/K_p$ of non-compact type. Thus $G_p$ will be connected non-compact
simple split rank 1 Lie groups with finite center and $K_p\subset G_p$ will be
maximal compact subgroups. Let $\Gamma_p\subset G_p$ be discrete, co-compact,
torsion free subgroups.

Let $L_p:C^{\infty}(V(M_p))\mapsto C^{\infty}(V(M_p))$ be partial differential
operators acting on smooth sections of vector bundles $V(M_p)$.
Let $\chi_p$ be a finite-dimensional unitary representation of $\Gamma_{p}$,
let $\{\lambda_l(p)\}_{l=0}^{\infty}$ be the set of eigenvalues of the
second-order operator of Laplace type $L_p=-\Delta_{\Gamma_{p}}$ acting on
smooth sections of the vector bundle over $\Gamma_{p}\backslash X_{p}$ induced
by $\chi_{p}$, and let $n_l(\chi_p)$ denote the multiplicity of $\lambda_l(p)$.

We shall need further a suitable regularization of the determinant of an
elliptic differential operator, since the naive definition of the product of
eigenvalues gives rise to a badly divergent quantity. We shall make the choice
of zeta-function regularization (see (3.2)). The zeta function associated with
the operators ${\cal L}_p\equiv L_p+b_p$ have the form
$$
\zeta(s|{\cal L}_p)=\sum_ln_l(\chi_p)\{\lambda_l(p)+b_p\}^{-s}\mbox{,}
\eqno{(2.6)}
$$
here $b_p$ are arbitrary constants (endomorphisms of the vector bundle
$V(M_p)$), called in the physical literature the potential terms.
$\zeta(s|{\cal L}_p)$ is a well-defined analytic function for
$\Re s >\mbox{dim}(M_p)/2$, and can be analytically continued to a meromorphic
function on the complex plane ${\vec C}$, regular at $s=0$. One can define the
heat kernel of the elliptic operator ${\cal L}_p$ by
$$
\omega_{\Gamma_{p}}(t)\equiv\mbox{Tr}\left(e^{-t{\cal L}_p}\right)=\frac{-1}
{2\pi i}\mbox{Tr}\int_{{\cal C}_0}e^{-zt}(z-{\cal L}_p)^{-1}dz\mbox{,}
\eqno{(2.7)}
$$
where ${\cal C}_0$ is an arc in the complex plane ${\vec C}$. By standard
results in operator theory there exist $\epsilon,\delta >0$ such that for
$0<t<\delta$ the heat kernel expansion holds
$$
\omega_{\Gamma_{p}}(t)=\sum_{l=0}^{\infty}n_l(\chi_p)e^{-(\lambda_l(p)+b_p)t}
=\sum_{0\leq l\leq l_0} a_l
({\cal L}_p)t^{-l}+ O(t^\epsilon)\mbox{.}
\eqno{(2.8)}
$$
Eventually we would like also to take $b_p=0$, but for now we consider only
non-zero modes: $b_p+\lambda_l(p)>0$, $\forall l : \lambda_0(p)=0$, $b_p>0$.

\ss{The Explicit Form of the Zeta Function}

The following representations of $X_p$ up to local isomorphism can be chosen
$$
X_p=\left[ \begin{array}{ll}
SO_1(n,1)/SO(n)\,\,\,\,\,\,\,\,\,\,\,\,\,\,\,\,\,\,\,\,\,\,\,\,\,\,\,\,\,
\,\,\,\,\,\,\,\,\,(I) \\
SU(n,1)/U(n)\,\,\,\,\,\,\,\,\,\,\,\,\,\,\,\,\,\,\,\,\,\,\,\,\,\,\,\,\,\,\,
\,\,\,\,\,\,\,\,\,\,(II)\\
SP(n,1)/(SP(n)\otimes SP(1))\,\,\,\,\,(III)\\
F_{4(-20)}/Spin(9)\,\,\,\,\,\,\,\,\,\,\,\,\,\,\,\,\,\,\,\,\,\,\,\,\,\,\,\,
\,\,\,\,\,\,\,(IV)
\end{array} \right]
\mbox{,}
\eqno{(2.9)}
$$
where $n\geq 2$, and $F_{4(-20)}$ is the unique real form of $F_4$ (with
Dynkin diagram $\circ-\circ=\circ-\circ$) for which the character
$(\mbox{dim}X - \mbox{dim}K)$ assumes the value $(-20)$ \cite{helg62}.
We assume that if $G_{1}$ or $G_{2}=SO(m,1)$ or $SU(q,1)$ then $m$ is even and
$q$ is odd.

We study the zeta function
$$
\zeta(s|\bigoplus_{p}{\cal L}_p)=\zeta_{\Gamma_1\backslash X_1\bigotimes\Gamma_2
\backslash X_2}(s;b_1b_2)=\frac{1}{\Gamma(s)}\int_0^\infty
\prod_{p}\omega_{\Gamma_p}(t)
t^{s-1}dt\mbox{,} \hspace{0.2cm}\Re s>\frac{d_1+d_2}{2}
\mbox{,}
\eqno{(2.10)}
$$
where $d_p=\mbox{dim}X_p$.
The suitable Harish-Chandra-Plancherel measure is given as follows:
$$
|C_p(r)|^{-2}=C_{G_p}\pi rP_p(r)\tanh \left(a(G_p)r\right)=
C_{G_p}\pi\sum_{l=0}^{\frac{d_p}{2}-1}a_{2l}(p)r^{2l+1}\tanh \left(a(G_p)r
\right)\mbox{,}
\eqno{(2.11)}
$$
where
$$
a(G_p)=\left[ \begin{array}{ll}
\pi \hspace{0.5cm}\mbox{for $G_p=SO_1(2n,1)$}\\
\frac{\pi}{2} \hspace{0.5cm}\mbox{for $G_p=SU(q,1),\,\,\,\,\,\,\,\, q$ odd}\\
\hspace{0.8cm}\mbox{or $G_p=SP(m,1),\,\,\,\,\,\, F_{4(-20)}$}
\end{array} \right]
\mbox{,}
\eqno{(2.12)}
$$
while $C_{G_p}$ is some constant depending on $G$, and where the $P_p(r)$ are
even polynomials (with suitable coefficients $a_{2l}(p)$) of degree $d-2$ for
$G\neq SO(2n+1,1)$, and of degree $d-1=2n$ for $G=SO_1(2n+1,1)$
\cite{byts96-266-1,will97-38-796}.

The explicit construction during the proof of Eq. (A.14) (of the Appendix A)
gives a little more, namely

\begin{theorem}

The function $\zeta(s|\bigoplus {\cal L}_p)$ admits an explicit meromorphic
continuation to ${\bf C}$ with at most a simple pole at $s=1,2,...,
\frac{d_1+d_2}{2}$. In particular on the domain $\Re s<1$,
$$
\zeta(s|\bigoplus_{p}{\cal L}_p)=\frac{\pi^2}{2}\prod_k^2V_ka(G_k)C_{G_k}
\sum_{j=0}^{\frac{d_1}{2}
-1}\sum_{l=0}^j\sum_{\mu=0}^{\frac{d_2}{2}-1}\sum_{\nu=0}^{\mu}
\frac{a_{2j}(1)a_{2\mu}(2)j!\mu!}{(j-l)!(\mu-\nu)!}
$$
$$
\times\frac{\int_0^{\infty} r^{2(j-l)}\mbox{sech}^2(a
(G_1)r)K_{\mu-\nu}(s-l-\nu-1;r^2+B,a(G_1))dr}
{(s-1)(s-2)...(s-(l+1))(s-(l+2))...(s-(l+1+\nu+1))}
$$
$$
+C_{G_1}V_1\sin (\pi s)\sum_{j=0}^{\frac{d_1}{2}-1}a_{2j}(1)\int_{\bf R}dr
r^{2j+1}\mbox{tanh}(a(G_1)r)
$$
$$
\times\left[\int_0^{\infty}dt\Psi_{\Gamma_2}\left(
\rho_0(2)+t+\sqrt{r^2+B};\chi_2\right)
\left(2t\sqrt{r^2+B}+t^2\right)^{-s}\right]
$$
$$
+C_{G_2}V_2\sin (\pi s)\sum_{j=0}^{\frac{d_2}{2}-1}a_{2j}(2)\int_{\bf R}dr
r^{2j+1}\mbox{tanh}(a(G_2)r)
$$
$$
\times\left[\int_0^{\infty}dt\psi_{\Gamma_1}\left(
\rho_0(1)+t+\sqrt{r^2+B};\chi_1\right)
\left(2t\sqrt{r^2+B}+t^2\right)^{-s}\right]
$$
$$
+\frac{1}{2\pi^3i\Gamma(s)}\int_{\Re z=\epsilon}dz[\sin \pi(z+\frac{s}{2})]
[\sin \pi(\frac{s}{2}-z)]\Gamma(z+\frac{s}{2})\Gamma(\frac{s}{2}-z)
$$
$$
\times\prod_k^2\left[\int_0^{\infty}dt\Psi_{\Gamma_k}\left(\rho_0(k)+t+
B_k^{\frac{1}{2}};\chi_1\right)\left(2tB_k^{\frac{1}{2}}+t^2\right)^
{-y_k(s,z)}\right]
\mbox{,}
\eqno{(2.13)}
$$
for any $-\frac{1}{2}\leq\epsilon\leq\frac{1}{2}$. All of the integrals are
entire functions of $s$.
\end{theorem}

The simplest case is, for example, $G_1=G_2=G=SO_1(2,l)\simeq SL(2,R)$; besides
$X_1=X_2=H^2$ is a two-dimensional real hyperbolic space. Then we have
$\rho_0^2=\rho_0^2(1)+\rho_0^2(2)=\frac{1}{4}+\frac{1}{4}=\frac{1}{2}$,
$a_{20}=1$, $C_G=1$, $a(G)=\pi$, $\Gamma_1=\Gamma_2=\Gamma$, and finally
$|C(r)|^{-2}=\pi r\tanh(\pi r)$. Using the Eq. (2.13) of the Theorem 1 for
$\Re s<1$ we have
$$
\zeta(s|\bigoplus_{p}{\cal L}_p)=\frac{\pi^3V_1^2}{2(s-1)(s-2)}\int_0^
{\infty}dr\mbox{sech}^2(\pi r)K_0\left(s-2;r^2+B,\pi\right)
$$
$$
+2V_1\sin (\pi s)\int_{R}drr\mbox{tanh}(\pi r)
\int_0^{\infty}dt\psi_{\Gamma}
\left(\frac{1}{2}+t+\sqrt{r^2+B};\chi\right)
\left(2t\sqrt{r^2+B}+t^2\right)^{-s}
$$
$$
+\frac{1}{2\pi^3i\Gamma(s)}\int_{\Re z=\epsilon}dz[\sin \pi(z+\frac{s}{2})]
[\sin \pi(\frac{s}{2}-z)]\Gamma(z+\frac{s}{2})\Gamma(\frac{s}{2}-z)
$$
$$
\times\int_0^{\infty}dt\psi_{\Gamma}\left(\frac{1}{2}+t+B_1^{\frac{1}{2}}
;\chi\right)\left(2tB_1^{\frac{1}{2}}+t^2\right)^{-y_1(s;z)}
$$
$$
\times\int_0^{\infty}dt\psi_{\Gamma}\left(\frac{1}{2}+t+B_2^{\frac{1}{2}}
;\chi\right)\left(2tB_2^{\frac{1}{2}}+t^2\right)^{-y_2(s;z)}
\mbox{,}
\eqno{(2.14)}
$$
where $B=\frac{1}{2}+b_1+b_2$ and $B_p=\frac{1}{4}+b_p\,\,\,(p=1,2)$.

\s{The Multiplicative Anomaly and Associated One-Loop Contributions}

In this section the product of the second-order operators of Laplace type
$\bigotimes{\cal L}_p$,\,\,$p=1,2$ will be considered. We are interested in
multiplicative properties of determinants, the multiplicative anomaly
\cite{kass89,kont94u-56,kont94u-40}, related with one-loop approximation in
quantum field theory. This approximation can be given in terms of the
multiplicative anomaly $F({\cal L}_1,{\cal L}_2)$, which has the form
$$
F({\cal L}_1,{\cal L}_2)=\mbox{det}_\zeta[\bigotimes_{p}{\cal L}_p]
[\mbox{det}_\zeta({\cal L}_1)\mbox{det}_\zeta({\cal L}_2)]^{-1}
\mbox{,}
\eqno{(3.1)}
$$
where we assume a $\zeta$-regularization of determinants, i.e.
$$
\mbox{det}_\zeta({\cal L}_p)\stackrel{def}{=}\exp\left(-\frac{\partial}
{\partial s}\zeta(s=0|{\cal L}_p)\right)\mbox{.}
\eqno{(3.2)}
$$
Generally speaking, if a multiplicative anomaly related to elliptic operators
is nonvanishing then the relation $\mbox{log}\mbox{det}(\bigotimes {\cal L}_p)=
\mbox{Tr}\mbox{log}(\bigotimes {\cal L}_p)$ does not hold.

The operator product $\bigotimes{\cal L}_p$ can arise in higher-derivative
field theories. Note also that the partition function $Z$, associated with the
product of two elliptic differential operators, for the simplest $O(2)$
invariant model of self-interacting charged fields in ${\bf R}^4$
\cite{bens91-44-2480} has been analyzed recently in Ref. \cite{eliz97u-60}:
$$
\mbox{log}Z\propto-\mbox{log}\mbox{det}\left(\bigotimes{\cal L}_p\right)
\mbox{.}
\eqno{(3.3)}
$$

\begin{theorem}
Given the notation and results of Eqs. (B.11), (B.17), (B.18) and Theorem 4
of Appendix B the explicit formula for the multiplicative anomaly is
$$
{\cal A}({\cal  L}_1,{\cal L}_2)=A\sum_{j=0}^{\frac{d}{2}-1}\frac{a_{2j}
(-1)^j}{2}\left\{\frac{j}{2}(B_1-B_2)^2B_2^{j-1}+\frac{j(j-1)}{4}
(B_1-B_2)^3B_2^{j-2}\right.
$$
$$
\left.+\sum_{p=3}^j\frac{j!}{(p+1)p!(j-p)!}\left[\frac{1}{p}+\frac{1}{p-1}
+\sum_{q=1}^{p-2}\frac{1}{p-q-1}\right](B_1-B_2)^{p+1}B_2^{j-p}\right\}
\mbox{.}
\eqno{(3.4)}
$$
\end{theorem}

In a special case, namely for $d=2$, Theorem 2 gives ${\cal A}({\cal L}_1,
{\cal L}_2)=0$. Finally for any odd $d$ the multiplicative anomaly is zero.
This statement follows from the general theory of Laplace type operators (see,
for example, Ref. \cite{eliz97u-60}).

\s{The Residue Formula and the Multiplicative Anomaly}

The value of $F({\cal L}_1,{\cal L}_2)$ can be expressed by means of the
non-commutative Wodzicki residue \cite{wodz87}.
Let ${\cal O}_p, \,\,\,p=1,2$ be invertible elliptic PDOs of real non-zero
orders $\al$ and $\be$ such that $\al+\be\neq 0$. Even if the zeta functions
for operators ${\cal O}_1, {\cal O}_2$ and ${\cal O}_1\bigotimes{\cal O}_2$ are
well defined and if their principal symbols obey the Agmon-Nirenberg condition
(with appropriate spectra cuts) one has in general that
$F({\cal O}_1,{\cal O}_2)\neq 1$. For such invertivble elliptic operators the
formula for the anomaly of commuting operators holds \cite{eliz97u-60}:
$$
{\cal A}({\cal O}_1,{\cal O}_2)={\cal A}({\cal O}_2,{\cal O}_1)=
\mbox{log}(F({\cal O}_1,{\cal O}_2))=
\left(2\al\be(\al+\be)\right)^{-1}\mbox{res}\left[(\mbox{log}({\cal O}_1^{\be}
\bigotimes{\cal O}_2^{-\al}))^2\right]\mbox{.}
\eqno{(4.1)}
$$
More general formulae have been derived in Refs. \cite{kont94u-56,kont94u-40}.
Furthermore the anomaly can be iterated consistently. Indeed, using Eqs. (3.2)
and (4.1) we have
$$
\,\,\,{\cal A}({\cal O}_1,{\cal O}_2)= \zeta'(0|{\cal O}_1{\cal O}_2)
-\zeta'(0|{\cal O}_1)-\zeta'(0|{\cal O}_2)\mbox{,}
$$
$$
\,\,\,\,\,\,\,\,{\cal A}({\cal O}_1,{\cal O}_2,{\cal O}_3)=
\zeta'(0|\bigotimes_{j}^3{\cal O}_j)
-\sum_j^3\zeta'(0|{\cal O}_j)-{\cal A}({\cal O}_1,{\cal O}_2)\mbox{,}
$$
$$
. \qquad . \qquad . \qquad  . \qquad . \qquad . \qquad . \qquad . \qquad .
\qquad . \qquad .
$$
$$
{\cal A}({\cal O}_1,{\cal O}_2,...,{\cal O}_n)= \zeta'(0|\bigotimes_{j}^n
{\cal O}_j)
-\sum_j^n\zeta'(0|{\cal O}_j)-{\cal A}({\cal O}_1,{\cal O}_2)
$$
$$
\qquad\qquad\qquad\qquad-{\cal A}({\cal O}_1,{\cal O}_2,{\cal O}_3)...
-{\cal A}({\cal O}_1,{\cal O}_2,...,{\cal O}_{n-1})\mbox{.}
\eqno{(4.2)}
$$
In particular, for $n=2$ and ${\cal O}_p\equiv{\cal L}_p$ the explicit form
of anomaly is given by the Eqs. (B.18) (of the Appendix B) and (3.4).

We note that for the four-dimensional space with $G=SO_1(4,1)$, one derives
from Theorem 2 the result
$$
{\cal A}({\cal L}_1,{\cal L}_2)=-A_G(b_1-b_2)^2,\,\,\,\,\,\,\,\,\, d=4\mbox{,}
\eqno{(4.3)}
$$
which also follows from Wodzicki's formula (4.1), where $A_G=\frac{1}{4}Aa_2$.

\s{Conclusions}

In this paper the global additive and multiplicative properties of Laplace
type operators and related zeta functions has been studied. We have considered
product structures on Einstein manifolds, especially on an irreducible rank 1
symmetric spaces.

In fact the explicit form of the zeta function on product spaces (Theorem 1)
is derived. As an example the zeta function associated with the
Kr{\"o}necker sum of Laplacians acting on two-dimensional real hyperbolic
spaces is calculated.

We have obtained also the explicit formula for the multiplicative anomaly, in
the main theorem, Theorem 2. It has been shown that the anomaly is equal to
zero for $d=2$ and for the odd dimensional cases; this result is in agreement
with the calculation given in Ref. \cite{eliz97u-60}. We have preferred to
limit ourselves here to discuss in detail various particular cases and
emphasize the general picture. It seems to us that the explicit results for the
anomaly (3.4), (4.3) are not only interesting as mathematical results but are
of physical interest, in view of future applications to concrete problems in
field theory and gravity, both at classical and quantum level. Note also that
spectral properties of products of differential operators related to higher
spin fields might differ in principal from the properties considered in this
paper; we hope return to this problem elsewhere.

\ack{ We thank Profs. E. Elizalde, L. Vanzo and S. Zerbini for useful
discussions. The research of A.A.B.~was supported in part by Russian
Foundation for Fundamental Research grant No.~95-02-03568-a and by Russian
Universities grant No.~95-0-6.4-1.}

\appendix

\s{Zeta Functions on Product of Rank 1 Symmetric Spaces}

In this Appendix we consider the trace formula for the partition function and
zeta function associated with the product of symmetric spaces.
Let the data $(G,K,\Ga)$ be as in Sect. 2, therefore $G$ being one of the
four groups of Eq. (2.9). The trace formula holds
\cite{wall76-82-171,will90-242}
$$
\omega_{\Ga}(t;b,\chi)=V
\int_{\bf R}dre^{-(r^2+b+\rho_0^2)t}|C(r)|^{-2}+\theta_{\Ga}(t;b,\chi)\mbox{,}
\eqno{(A.1)}
$$
where, by definition,
$$
V\stackrel{def}{=}\frac{1}{4\pi}\chi(1)\mbox{vol}(\Ga\backslash G)
\mbox{,}
\eqno{(A.2)}
$$
where $\chi$ is a finite-dimensional unitary representation (or a character) of
$\Ga$, and the number $\rho_0$ is associated with the positive restricted
(real) roots of $G$ (with multiplicity) with respect to a nilpotent factor $N$
of $G$ in an Iwasawa decomposition $G=KAN$. One has $\rho_0=(n-1)/2, n, 2n+1,
11$ in the cases $(I)-(IV)$ respectively in Eq. (2.9). Finally the function
$\theta_{\Ga}(t;b,\chi)$ is defined as follows
$$
\theta_{\Ga}(t;b,\chi)\stackrel{def}{=}\frac{1}{\sqrt{4\pi t}}\sum_{\ga\in C_
\Ga-\{1\}}\chi(\ga)t_\ga j(\ga)^{-1}C(\ga)e^{-(bt+\rho_0^2t+t_\ga^2/(4t))}
\mbox{,}
\eqno{(A.3)}
$$
for a function $C(\ga)$,\,\, $\ga\in\Ga$, defined on $\Ga-\{1\}$ by
$$
C(\ga)\stackrel{def}=e^{-\rho_0t_\ga}|\mbox{det}_{n_0}\left(\mbox{Ad}(m_\ga
e^{t_\ga H_0})^{-1}-1\right)|^{-1}\mbox{.}
\eqno{(A.4)}
$$

The notation used in Eqs. (A.3) and (A.4) is the following. Let $a_0, n_0$
denote the Lie algebras of $A, N$. Since the rank of $G$ is 1,
$\dim a_0=1$ by definition, say $a_0={\bf R}H_0$ for a suitable basis vector
$H_0$. One can normalize the choice of $H_0$ by $\beta(H_0)=1$, where
$\beta: a_0\mapsto{\bf R}$ is the positive root which defines $n_0$; for more
detail see Ref. \cite{will97-38-796}. Since $\Ga$ is torsion free, each
$\ga\in\Ga-\{1\}$ can be represented uniquely as some power of a primitive
element $\delta:\ga=\delta^{j(\ga)}$ where $j(\ga)\geq 1$ is an integer and
$\delta$ cannot be written as $\ga_1^j$ for $\ga_1\in \Ga$, \,\,$j>1$ an
integer. Taking $\ga\in\Ga$, $\ga\neq 1$, one can find $t_\ga>0$
and $m_\ga\in K$ satisfying $m_\ga a=am_\ga$ for every $a\in A$ such that $\ga$
is $G$ conjugate to $m_\ga\exp(t_\ga H_0)$, namely for some
$g\in G, \,g\ga g^{-1}=m_\ga\exp(t_\ga H_0)$. For $\mbox{Ad}$ denoting the
adjoint representation of $G$ on its complexified Lie algebra, one can compute
$t_\ga$ as follows \cite{wall76-82-171}
$$
e^{t_\ga}=\mbox{max}\{|c||c= \mbox{an eigenvalue of}\,\, \mbox{Ad}(\ga)\}
\mbox{,}
\eqno{(A.5)}
$$
in case $G=SO_1(m,1)$, with $|c|$ replaced by $|c|^{1/2}$ in the other cases
of Eq. (2.9).

\ss{Zero Modes}

Let us start with the zero modes case, i.e. $b=0$. It can be shown \cite{will96}
that the Mellin transform of $\theta_\Ga(t,b,\chi)$ at $b=0$,
$$
\hat{\theta}_\Ga(s;0,\chi)\stackrel{def}{=}\int_0^{\infty}dt\theta_\Ga
(t;0,\chi)t^{s-1}\mbox{,}
\eqno{(A.6)}
$$
is a holomorphic function on the domain $\Re s<0$. Then using the result of
Refs. \cite{byts96-266-1,will97-38-796} one can obtain on $\Re s<0$,
$$
\hat{\theta}_{\Ga}(s;0,\chi) = \sum_{\ga\in C_\Ga-\{1\}}
\chi(\ga)t_\ga j(\ga)^{-1}C(\ga)\int_0^{\infty}dt
\frac{e^{-(\rho_0^2t+t_\ga^2/(4t))}}{\sqrt{4\pi t}}t^{s-1}
$$
$$
\qquad\qquad\quad=\frac{(2\rho_0)^{\frac{1}{2}-s}}{\sqrt{\pi}}
\sum_{\ga\in C_\Ga-\{1\}}
\chi(\ga)t_\ga j(\ga)^{-1}C(\ga)t_\ga^{s+\frac{1}{2}}
K_{\frac{1}{2}-s}(t_\ga\rho)\mbox{,}
\eqno{(A.7)}
$$
where $K_\nu(s)$ is the modified Bessel function, and finally
$$
\hat{\theta}_\Ga(s;0,\chi)=\frac{\sin (\pi s)}{\pi}\Ga(s)\int_0^{\infty}dt
\psi_\Ga(t+2\rho_0;\chi)(2\rho_0t+t^2)^{-s}\mbox{.}
\eqno{(A.8)}
$$
Here $\psi_\Ga(s;\chi)\equiv d(\mbox{log}Z_\Ga(s;\chi))/ds$,\,\,\, and
$Z_\Ga(s;\chi)$ is a meromorphic suitably normalized Selberg zeta function
\cite{selb56-20-47,frie77-10-133,gang77-21-1,gang80-78-1,scot80-253-177,
waka85-15-235,will90-242,will92-105-163,byts96-266-1}.

\ss{A Meromorphic Continuation}

For $B_p\stackrel{def}{=}b_p+\rho_0^2(p)\,\,(p=1,2),\,\,\,y_p(s;z)\stackrel{def}
{=}\frac{s}{2}+(-1)^{p-1}z$, the following proposition holds:

\begin{proposition}

The integral of the product of the functions
$\hat{\theta}_{\Ga_p}(y_p(s;z),b_p,\chi_p)$ is an entire function of $s$ and
allows the form
$$
\int_0^{\infty}dt\prod_p\theta_{\Ga_p}
(t;b_p,\chi_p)t^{s-1}=
\frac{1}{2\pi i}\int_{\Re z=c}dz\prod_p\hat{\theta}_{\Ga_p}
(y_p(s,z); b_p, \chi_p)
$$
$$
=\frac{1}{2i\pi^3}\int_{\Re z=c}dz\prod_l^2\Ga(y_l(s;z))
[\sin \pi(y_l(s,z))]
$$
$$
\times\prod_p\int_0^{\infty}dt\psi_{\Ga_p}(\rho_0(p)+t+B_p^{\frac{1}{2}};\chi_p)
(2B_p^{\frac{1}{2}}t+t^2)^{-y_p(s;z)}
\mbox{,}
\eqno{(A.9)}
$$
for $c\in{\bf R}$ with $\Re \frac{s}{2}<c<-\Re\frac{s}{2}$, and $b_p\geq 0$
with at least $b_1>0$ or $b_2>0$.
\end{proposition}
The proposition is a consequence of the first integral transformation of
Eq. (A.9) as an integral in the complex plane with the help of the
Mellin-Parseval identity
$$
\int_0^{\infty}dth(t)g(t)=\frac{1}{2\pi i}\int_{\Re z=c}dz\hat{h}(z)\hat{g}
(1-z),
\,\,\,\,c\in{\bf R}
\mbox{.}
\eqno{(A.10)}
$$
To this end one can choose $h(t)=\theta_{\Ga_1}(t;b_1,\chi_1)t^{s/2}, \,\,\,
g(t)=\theta_{\Ga_1}(t;b_2,\chi_2)t^{s/2-1}$.

We shall also need the $B\stackrel{def}{=}\rho_0^2(1)+\rho_0^2(2)+b_1+b_2$,
the function
$$
{\cal K}_n(s;\delta,a)\stackrel{def}{=}\int_{\bf R}\frac{drr^{2n}\mbox{sech}^2
(ar)}{(\delta+r^2)^s}\mbox{,}
\eqno{(A.11)}
$$
defined for $a,\delta>0,\,\,n\in {\cal N}\,\,s\in{\bf C}$, and the functions
$$
f_l(t;a,G)\stackrel{def}{=}\int_0^{\infty}drr^{2l+1}e^{-(r^2+a)t}\tanh(a(G)r)
\mbox{,}
\eqno{(A.12)}
$$
$$
H_\Ga^{(l)}(s;a,b,\chi,G)\stackrel{def}{=}\int_0^{\infty}dtf_l(t;a,G)\theta_
\Ga(t;b,\chi)t^{s-1}\mbox{.}
\eqno{(A.13)}
$$
All the functions (A.11) - (A.13) are entire functions of variable $s$.

\begin{proposition}
Suppose $G_p\neq SO_1(m,1),\,\,\, SU(q,1)$ with $m$ odd and $q$ even. Then for
$Re s>\frac{d_1}{2}+\frac{d_2}{2}$,
$$
\frac{1}{\Ga(s)}\int_0^{\infty}dt\prod_p\omega_{\Ga_p}(t,b_p,\chi_p)t^{s-1}=
\frac{\pi^2}{2}\prod_k^2V_ka(G_k)C_{G_k}
\sum_{j=0}^{\frac{d_1}{2}
-1}\sum_{l=0}^j\sum_{\mu=0}^{\frac{d_2}{2}-1}\sum_{\nu=0}^{\mu}
\frac{a_{2j}(1)a_{2\mu}(2)j!\mu!}{(j-l)!(\mu-\nu)!}
$$
$$
\times\frac{\int_0^{\infty} drr^{2(j-l)}\mbox{sech}^2(a
(G_1)r){\cal K}_{\mu-\nu}(s-l-\nu-1;r^2+B,a(G_1))dr}
{(s-1)(s-2)...(s-(l+1))(s-(l+2))...(s-(l+1+\nu+1))}
$$
$$
+\frac{2\pi}{\Ga(s)}\left\{
V_1C_{G_1}\sum_{j=0}^{\frac{d_1}{2}-1}a_{2j}(1)H_{\Ga_2}^{(j)}
(s;b_1+B_1,b_2,\chi_2,G_1)\right.
$$
$$
\left.+V_2C_{G_2}\sum_{j=0}^{\frac{d_2}{2}-1}a_{2j}(2)H_{\Ga_1}^{(j)}
(s;b_2+B_2,b_1,\chi_1,G_2)\right\}
$$
$$
+\frac{1}{\Ga(s)}\int_0^{\infty}dt\prod_k^2\theta_{\Ga_k}(t,b_k,\chi_k)t^{s-1}
\mbox{.}
\eqno{(A.14)}
$$

\end{proposition}

The formulae (A.9), (A.14) give the meromorphic continuation of the zeta
function (2.10).

\s{The Zeta Function of the Product of Laplace-Type Operators}

The spectral zeta function associated with the product $\bigotimes {\cal L}_p$
has the form
$$
\zeta(s|\bigotimes_p{\cal L}_p)=\sum_{j\geq 0}n_j\prod_p^2(\lambda_j+b_p)^{-s}
\mbox{.}
\eqno{(B.1)}
$$
We shall always assume that $b_1\neq b_2$, say $b_1>b_2$. If $b_1=b_2$ then
$\zeta(s|\bigotimes{\cal L}_p)=\zeta(2s|{\cal L})$ is a well-known function.
For $b_1,b_2\in{\bf R}$, set $b_{+}\stackrel{def}{=}(b_1+b_2)/2,\,\, b_{-}
\stackrel{def}{=}(b_1-b_2)/2$, thus $b_1=b_{+}+b_{-}$ and $b_2=b_{+}-b_{-}$.

\begin{remark}

The spectral zeta function can be written as follows
$$
\zeta(s|\bigotimes_{p}{\cal L}_p)=(2b_{-})^{\frac{1}{2}-s}\frac{\sqrt{\pi}}{\Ga(s)}
\int_0^{\infty}dt\omega_\Ga(t,b_{+})I_{s-\frac{1}{2}}(b_{-}t)
\mbox{,}
\eqno{(B.2)}
$$
where the integral converges absolutely for $Re s>\frac{d}{4}\,\,
(d=\mbox{dim}(G/K))$.
\end{remark}
This formula is a main starting point to study the zeta function. It expresses
$\zeta(s|\bigotimes{\cal L}_p)$ in terms of the Bessel function
$I_{s-\frac{1}{2}}(b_-t)$ and $\omega_\Ga(t,b_{+})$, where the trace formula
applies to $\omega_\Ga(t,b_{+})$. Thus we can study the zeta function by the
trace formula. One can use the trace formula for $\omega_\Ga(t,b_{+})$ (A.1);
as a result
$$
\zeta(s|\bigotimes_{p}{\cal L}_p)=(2b_{-})^{\frac{1}{2}-s}\frac{\sqrt{\pi}}
{\Ga(s)}\int_0^{\infty}dt\left[\frac{\chi(1)\mbox{vol}(\Ga\backslash G)}{4\pi}
\int_{\bf R}dre^{-(r^2+b_++\rho_0^2)t}|C(r)|^{-2}+
\theta_\Ga(t)\right]I_{s-\frac{1}{2}}t^{s-\frac{1}{2}}\mbox{.}
\eqno{(B.3)}
$$
Then for $\Re s>0$ Fubini's theorem gives
$$
\int_0^{\infty}dt\frac{\chi(1)\mbox{vol}(\Ga\backslash G)}{4\pi}
\int_{\bf R}dre^{-(r^2+b_++\rho_0^2)t}|C(r)|^{-2}I_{s-\frac{1}{2}}
t^{s-\frac{1}{2}}=
$$
$$
(2b_{-})^{\frac{1}{2}-s}\frac{\chi(1)\mbox{vol}(\Ga
\backslash G)\Ga(s)}{4\pi^{3/2}}\int_{\bf R}dr|C(r)|^{-2}\prod_p(r^2+B_p)^{-s}
\mbox{.}
\eqno{(B.4)}
$$
In order to analyze the last integral in Eq. (B.4) (for the possibility of
a meromorphic continuation) it is useful to rewrite the function $|C(r)|^{-2}$
(see Eq. (2.11)), using the identity $\mbox{tanh}(ar)\equiv 1-2(1+e^{2ar})^
{-1}$. Then one can calculate a suitable integral in terms of the hypergeometric
function $F(\al,\be;\ga;z)$, namely
$$
\int_0^{\infty}drr^{2j+1}\prod_{p}(r^2+B_p)^{-s}=
\frac{\sqrt{\pi}\Ga(2s-j-1)j!}{2^{2s}\Ga(s)\Ga(s+\frac{1}{2})}B_1^{-s}
B_2^{j+1-s}\left(\frac{2B_1}{B_1+B_2}\right)^{j+1}
$$
$$
\qquad\qquad\qquad\qquad\quad \times F\left(\frac{j+1}{2},\frac{j+2}{2};
s+\frac{1}{2};\left(\frac{B_1-B_2}{B_1+B_2}\right)^2\right)\mbox{,}
\eqno{(B.5)}
$$
which is holomorphic function on $\Re s> (j+1)/2$ and admits a meromorphic
continuation to ${\bf C}$ with only simple poles at points $s=(j+1-n)/2,\,\,
n\in{\cal N}$. Let us introduce the function
$$
E_j(s)\stackrel{def}{=}2\int_0^{\infty}\frac{drr^{2j+1}}
{(1+e^{2a(G)r})}\prod_{p}(r^2+B_p)^{-s}\mbox{,}
\eqno{(B.6)}
$$
which is an entire function of $s$. Then a consequence of Eqs. (B.2) - (B.6)
is

\begin{theorem}

For $Re s>\frac{d}{4}$ the explicit meromorphic continuation holds:
$$
\zeta(s|\bigotimes_{p}{\cal L}_p)={\cal F}(s)-2A\sum_{j=0}^{\frac{d}{2}-1}
a_{2j}E_j(s)+{\cal I}(s)\mbox{,}
\eqno{(B.7)}
$$
where
$$
A\stackrel{def}{=}\frac{1}{4}\chi(1) {\rm vol}(\Ga \backslash G)C_G\mbox{,}
\eqno{(B.8)}
$$
$$
{\cal F}(s)\stackrel{def}{=}A(B_1B_2)^{-s}\sum_{j=0}^{\frac{d}{2}-1}
\frac{a_{2j}j!\left(\frac{2B_1B_2}{B_1+B_2}\right)^{j+1}F\left(\frac{j+1}{2},
\frac{j+2}{2};s+\frac{1}{2};\left(\frac{B_1-B_2}{B_1+B_2}\right)^2\right)}
{(2s-1)(2s-2)...(2s-(j+1))}\mbox{,}
\eqno{(B.9)}
$$
$$
{\cal I}(s)\stackrel{def}{=}(2b_-)^{\frac{1}{2}-s}\frac{\sqrt{\pi}}{\Ga(s)}
\int_0^{\infty}dt\theta_\Ga(t,b_+)I_{s-\frac{1}{2}}(b_-t)t^{s-\frac{1}{2}}
\mbox{.}
\eqno{(B.10)}
$$
\end{theorem}

The goal now is to compute the derivative of the zeta function at $s=0$. Thus we
have
$$
\zeta'(0|\bigotimes_{p}{\cal L}_p) = A\sum_{j=0}^{\frac{d}{2}-1}a_{2j}\sum_l^4
{\cal E}_l\mbox{,}
\eqno{(B.11)}
$$
where
$$
{\cal E}_1=j!(B_1^{j+1}+B_2^{j+1})\sum_{k=0}^j\frac{(-1)^{k+1}}{k!(j-k)!
(j+1-k)!}\mbox{,}
\eqno{(B.12)}
$$
$$
{\cal E}_2=B_2^{j+1}\left(\frac{B_1-B_2}{2B_1}\right)\frac{(-1)^j}{(j+1)!}
\sum_{k=1}^{\infty}\frac{(j+k+1)!}{(k+1)!}\sigma_k\left(\frac{B_1-B_2}{B_1}
\right)^k\mbox{,}
\eqno{(B.13)}
$$
$$
{\cal E}_3=\mbox{log}(B_1B_2)\frac{(-1)^j}{2(j+1)}(B_1^{j+1}+B_2^{j+1})
-4\int_0^{\infty}\frac{drr^{2j+1}\mbox{log}\left(\frac{r^2+B_1}{r^2+B_2}
\right)}{1+e^{2a(G)r}}\mbox{,}
\eqno{(B.14)}
$$
$$
{\cal E}_4\equiv{\cal I}'(s=0)=T_\Ga(0,b_1,\chi_1)+T_\Ga(0,b_2,\chi_2)\mbox{,}
\eqno{(B.15)}
$$
and
$$
T_\Ga(0,b_p,\chi_p)\stackrel{def}{=}\int_0^{\infty}dt\theta_\Ga(t,b_p)t^{-1},
\,\,\,\,\,\,\,\, \sigma_k\stackrel{def}{=}\sum_{k=1}^n\frac{1}{k}
\mbox{.}
\eqno{(B.16)}
$$
Similary using results from Ref. \cite{will96} one can show
$$
\zeta'(0|{\cal L}_p)= A\sum_{j=0}^{\frac{d}{2}}a_{2j}\left\{j!B_p^{j+1}\left[
\sum_{k=0}^j\frac{(-1)^{k+1}}{k!(j-k)!(j+1-k)^2}+\frac{(-1)^j}{(j+1)!}
\log B_p\right]\right.
$$
$$
\left.+4\int_0^{\infty}\frac{drr^{2j+1}\log (r^2+B_p)}{1+e^{2a(G)r}}
\right\}+T_\Ga(0,b_p,\chi_p)\mbox{.}
\eqno{(B.17)}
$$
Then we can put these results together to compute the anomaly
$$
{\cal A}({\cal L}_1,{\cal L}_2)\stackrel{def}{=}\zeta'(0|\bigotimes{\cal L}_p)
-\zeta'(0|{\cal L}_1)-\zeta'(0|{\cal L}_2)\mbox{.}
\eqno{(B.18)}
$$
\begin{proposition}
A preliminary form of the multiplicative anomaly is

$$
{\cal A}({\cal  L}_1,{\cal L}_2)=A\sum_{j=0}^{\frac{d}{2}-1}a_{2j}B_2^{j+1}
\left(\frac{B_1-B_2}{2B_1}\right)\frac{(-1)^j}{(j+1)!}\sum_{k=1}^{\infty}
\sigma_k\left(\frac{B_1-B_2}{B_1}\right)^k\frac{(j+k+1)!}{(k+1)!}
$$
$$
+A\sum_p \log (B_p)\sum_{j=0}^{\frac{d}{2}-1}a_{2j}\frac{(-1)^{j+p}}
{2(j+1)}\left(B_1^{j+1}-B_2^{j+1}\right)\mbox{.}
\eqno{(B.19)}
$$

\end{proposition}

To obtain the final explicit form of the anomaly we must compute
the infinite series appearing in the statement of Proposition 3.
For $0<x<1$, $j=0,1,2,...$ let
$$
S_j(x)=\sum_{k=1}^{\infty}\frac{(j+k+1)!}{(k+1)!}\sigma_kx^k
\mbox{.}
\eqno{(B.20)}
$$
Thus in Proposition 3 we choose $x=(B_1-B_2)/B_1$. After a tedious calculation
one obtains the following result.
\begin{theorem}
The function $S_j(x)$ admits the representation
$$
(1-x)^{j+1}S_j(x)=P_j(x)+Q_j(x) \log (1-x)
\mbox{,}
\eqno{(B.21)}
$$
where $P_j(x)$ and $Q_j(x)$ are polynomials of degree $j$ given by
$$
\frac{P_j(x)}{(j+1)!}=\frac{jx}{2}(1-x)^{j-1}+\frac{j(j-1)}{4}x^2(1-x)^{j-2}
+\sum_{p=3}^j\frac{j!}{(p+1)!(j-p)!}
$$
$$
\times\left[\frac{1}{p}+\frac{1}{p-1}
+\sum_{q=1}^{p-2}\frac{1}{p-q-1}\right]x^p(1-x)^{j-p}\mbox{,}
\eqno{(B.22)}
$$
$$
\frac{xQ_j(x)}{(j+1)!}=-\frac{1}{j+1}\left[1-(1-x)^{j+1}\right]\mbox{.}
\eqno{(B.23)}
$$
\end{theorem}

\end{document}